\documentclass{aa}

\newcommand{\kms}{km\,s$^{-1}$}

\newcommand{\cmc}{cm$^{-3}$}
\newcommand{\scm}{cm$^{-2}$}
\newcommand{\ea}{et~al.~}
\newcommand{\cii}{C\,{\sc ii}}
\newcommand{\civ}{C\,{\sc iv}}

\newcommand{\ovi}{O\,{\sc vi}}
\newcommand{\nv}{N\,{\sc v}}

\usepackage{times}
\usepackage{graphics}
\begin{document}

\thesaurus{(
03.19.2; 
09.07.1; 
09.19.1; 
10.08.1; 
10.19.1; 
13.21.3; 
)}

\title{{\it Letter to the Editor}\\
ORFEUS\,II echelle spectra:\\
The scale height of interstellar \ovi\ in the halo}
\author{
H.\,Widmann\inst{1}\and
K.\,S.\,de Boer\inst{2}\and
P.\,Richter\inst{2}\and
G.\,Kr\"{a}mer\inst{1} \and
I.\,Appenzeller\inst{3} \and
J.\,Barnstedt\inst{1} \and
M.\,G\"{o}lz\inst{1} \and
M.\,Grewing\inst{4} \and
W.\,Gringel\inst{1} \and
H.\,Mandel\inst{3} \and
K.\,Werner\inst{1}}

\offprints{widmann@ait.physik.uni-tuebingen.de}

\institute{Institut f\"{u}r Astronomie und Astrophysik, Abt. Astronomie,
Universit\"{a}t T\"{u}bingen, Waldh\"{a}userstr. 64, D-72076 T\"{u}bingen, Germany \and
Sternwarte, Universit\"{a}t Bonn, Auf dem H\"{u}gel 71, D-53121 Bonn,
  Germany \and 
Landessternwarte Heidelberg, K\"{o}nigstuhl, D-69117 Heidelberg, Germany \and
Institut de Radio Astronomie Millim\'{e}trique (IRAM), 300 Rue de la Piscine, F-
38406 Saint
Martin d'H\`{e}res, France}

\titlerunning{The scale height of interstellar \ovi\ in the halo}

\date{Received $<$date$>$ / Accepted $<$date$>$ }

\maketitle

\begin{abstract} FUV high resolution spectra of 18 stars, particularly
  chosen to observe the interstellar medium (ISM), were obtained during
  the second ORFEUS-SPAS free flight space shuttle mission in December 1996. 
  Among these were 6 objects with a distance to the galactic plane larger
  than  1 \,kpc, one SMC and 4 LMC stars. 
  This selction of targets is part of the ORFEUS program to explore the 
  galactic halo. 
  As the most important tracer of the hot gas we analyzed the stronger
  component of the important \ovi\ doublet in all our ISM spectra.
  We found an average $N$(\ovi)$\times \sin|b|$ of $ \sim 3.5 \times
  10^{14}$\,cm$^{-2}$ on the lines of sight to the 4 LMC stars. 
  Assuming an exponential distribution of \ovi\ we 
  calculated  a midplane density  $n_\mathrm{0}$ of $ 2.07^{+0.26}_{-0.24}
  \times 10^{-8}$\cmc and a scale height $h_\mathrm{0}$ of
  $5.50^{+2.37}_{-2.09}$\,kpc.  


\end{abstract}

\section{Introduction}   
A hot galactic corona was first postulated by Spitzer (1956) as a
medium to confine the high velocity clouds discovered by M\"{u}nch (1952,
1957).  
M\"{u}nch found absorption in the Ca\,{\sc ii}\ lines at `high' velocities 
in spectra of stars at high galactic latitude. 

Almost all information on the hot gas distribution within or surrounding the
Galactic disk gathered since then derives from three different kinds of 
observation: the soft X-ray background at low and medium energies, the
high-stage ion populations (C\,{\sc iv}, N \,{\sc v} and \ovi) observed in
absorption in the UV and Far UV, and on the detected FUV and EUV
emission line backgrounds.  
Among the FUV absorption lines the \ovi\ ion contributes the most
important information to the understanding of the (hot) Galactic halo. 
It samples rather high temperature gas ($ \sim 10^{5.5}$\,K) with little
contamination expected from photoionised gas and it has a line strength
large enough that even nearby stars normally have detectable column
densities.    
With the Copernicus satellite the first detection of \ovi\ absorption
lines (Rogerson et al. 1973) at 1031.92\,\AA\ ($\log gf = -0.58$) and
1037.61\,\AA\ ($\log gf = -0.88$) was made 25 years ago.  
These data were analyzed and summarized in a series of papers by Jenkins 
(1978a, b, c). 
Excluding a few lines of sight on the grounds of atypically high $N$(\ovi), 
Jenkins found an average midplane density $n_\mathrm{0}$ of about
2.8\,$\times 10^{-8}\,$\,cm$^{-3}$ and a scale height $h_\mathrm{0}$ of 
300$^{+200}_{-150}$\,pc.   

Since the wavelength region of the \ovi\ resonance doublet is inaccessible for
IUE and HST the last available observations with reasonable wavelength 
resolution were made with the Berkeley spectrometer in Sept. 1993 during 
the ORFEUS I mission launched on the Space Shuttle Discovery. 
The ORFEUS-SPAS mission is discussed in detail in Grewing et al. (1991).
Hurwitz et al. (1995) present part of their spectrum of the SMC star NGC 346
No.1 near the \ovi\ resonance line and a revisit of the ORFEUS spectrum of 
PKS 2155-304. 
They derived from this sparse data an upper limit for $N$(\ovi) in the
galactic halo of 2.0 $\times 10^{14}\,$ cm$^{-2}$ toward NGC 346 No.1 and  
2.2 $\times 10^{14}\,$ cm$^{-2}$  towards PKS 2155-304. 
Hurwitz \& Boywer (1996b) analyzed 14 early type halo stars (obtained
with the Berkeley spectrometer) during the first mission.  
They derived a scale height for \ovi\ between about 80 pc and 600 pc if the 
midplane density is between 1.5 and 5 $\times 10^{-8}$ \cmc. 
These results were inconsistent with the comparativeley high column
densities of \nv\ reported by Sembach \& Savage (1992) and the \ovi/\nv\
ratios measured in disk stars and predicted by various theories (see 
Spitzer 1996). 
New observations, performed with the Berkeley spectrograph aboard the ORFEUS II
mission in Nov./Dec. 1996, yield to a $N$(\ovi) of $(7\pm2)\times 10^{14}\,$
cm$^{-2}$ toward the quasi-stellar object 3C 273 (Hurwitz et al. 1998).
 
Here we present first results from ORFEUS echelle spectra 
obtained during the mission of Nov./Dec. 1996. 
The strength of the absorption by the \ovi\ line at 1031.92 \AA\ has 
been determined and column densities have been calculated. 
The ensemble of data allows us to determine a substantially larger scale
height of \ovi\ in the galactic halo.   

\begin{table*}[tdb]
\caption[]{Basic properties of targets and parameters for \ovi}
\begin{tabular}{lrlcccccccccc}
\hline  
Object & $V$ &  Sp. T. & $l$ & $b$ & $z$  &$W_\lambda$ &
$N$(\ovi) & Error & S/N & Exposure & No. of  & Ref.$^{b}$  \\
       & [mag]     &         &    &     & [kpc] &[\AA]
       &\multicolumn{2}{c}{[$10^{14}$cm$^{-2}$]}  & & [ksec] & star& \\
\hline 
 HD 18100  &   8.46 & B1 V        &  217.9  & -62.7  &  2.67 & 0.177  &
 1.42  & 0.64  & 11 & 2.9  &   0 & 1,3    \\
 HD 49798  &   8.29 & sd O6 V    &  253.7  & -19.1  &  0.21 & 0.044  &    0.35  & 0.11  & 17 & 1.3  &   6 & 2   \\
 HD 77770  &   7.53 & B2.0 IV     &  169.3  &  41.9  &  0.77 & 0.086  &    0.69  & 0.14  &  9 & 1.9  &   8 & 3   \\
 HD 93521$^{a}$&   7.06 & O9.5 V     &  183.1  &  62.2  &  1.50 & 0.125  &    0.99  & 0.15  & 18 & 1.7  &  10 & 1,3    \\
 HD 93840  &   7.76 & B1 Ib       &  282.1  &  11.1  &  0.90 & 0.406  &    3.24  & 0.80  &  3 & 2.6  &   3 & 3   \\
 HD 116852 &   8.40 & O9 III     &  304.9  & -16.1  &  1.33 & 0.354  &    2.83 & 0.47  &  5 & 3.0  &  11 & 1   \\
 HD 146813 &   9.10 & B1.5 IV      &   85.7  &  43.8  &  1.82 & 0.114  &    0.91  & 0.37  &  5 & 1.4  &   7 & 3   \\
 HD 214930 &   7.38 & B2 VI      &   88.3  & -30.1  &  0.50 & 0.080  &    0.64  & 0.12  &  3 & 1.0  &   9 & 3   \\
 HD 217505 &   9.15 & B2 III     &  325.5  & -52.6  &  2.38 & 0.234  &    1.87  & 0.66  &  5 & 1.9  &   5 & 4   \\
 HD 36402  &  11.50 & OB+WC5, LMC&  277.8  & -33.0  & 30.0 & 0.740  &    5.91  & 2.46  &  2 & 1.6  &   4 & 5    \\
 HD 269546 &  11.30 & B3+WN3, LMC&  279.3  & -32.8  & 29.8 & 0.871  &    6.96  & 3.25  &  3 & 6.4  &   2 & 6  \\
 LH 10:3120&  12.80 & O5.5 V, LMC&  277.2  & -36.1  & 32.4 & 0.840  &    6.71  & 3.82  &  1 & 6.5  &  12 & 7   \\
 Sk $-$67\degr 166&  12.27 & O5e, LMC &  277.8  & -32.5  & 29.5 & 0.826  &    6.60  & 2.56  &  3 & 6.2  &  13 & 8  \\
 HD 5980   &  11.80 & OB+WN3, SMC&  302.1  & -44.9  & 45.9 & 0.193  &
 1.54  & 0.73  &  4 & 6.8  &   1 & 5  \\

\hline
\end{tabular}
                      
\noindent
$^{a}$ Jenkins (1978a) found an $N$(\ovi) of 7.24$ \times 10^{13}$cm$^{-2}$; 
$^{b}$ References: (1) Savage et al. (1997); (2) Jenkins \& Wallerstein
(1996); (3) Diplas \& Savage (1994); (4) Hurrwitz \& Boyer (1996b); (5)
Sembach \& Savage (1992); (6) Chu et. al (1994); (7) de Boer et al.(1998); 
(8) SIMBAD
\end{table*}

\section{Instrumentation and observation}
The ORFEUS 1m-telescope is equipped with two alternatively operating
spectrometers. 
The details about the telescope and the Echelle spectrometer
are discussed in Kr\"amer et al. (1990) and in Barnstedt \ea (1998), the
Berkeley spectrometer is described in Hurwitz \& Boywer (1996a).
The relevant properties of the Echelle spectrometer for the
measurements discussed are: a spectral range from 912\,\AA\ to
1410\,\AA, spectral resolution of $\lambda/\Delta\lambda \leq 10,000$ and an
effective collecting area of 1 cm$^{\rm 2}$.

Most of the stars on the ORFEUS\,II P.I. team target list were 
selected for interstellar and intergalactic medium research. 
We selected in particular high $|z|$ stars with large distances from the
galactic plane. 
Discussed in this paper are 9 Galactic and 5 Magellanic Cloud
objects; 6 of the Galactic targets have a $|z| >$ 1 kpc. 
Table 1 lists the targets and gives their basic properties.

\section{Data reduction}
\subsection{Continuum fitting and identification of other spectral lines} 

We binned the data in the echelle order of the \ovi\ line into elements
containing 7 pixel (of optical resolution), equivalent to 0.21 \AA\  and
fitted a 5th order polynominal to define the interstellar continuum. 
For the galactic targets, a straight line is a good approach to the
continuum of the narrow interstellar features (Fig. 1b).
A multi gaussian or even single gaussian fit (see HD 93521 in Fig. 1a) was
applied to the spectral structures, to define the equivalent width of 
the explored absorption.  

In the vicinity of the \ovi\ line at 1031.92 \AA\ absorption structures
due to H$_2$ Lyman P(3) at 1031.19 \AA\ and H$_2$ Lyman R(4) at 1032.35 \AA\ 
are present. 
However, these are well separable from the \ovi\ line. 
Further possible interference may arise from the R(0) 6-0 interstellar line
of HD at 1031.91\,\AA.
We determined the contributions from this feature by looking for the 7-0
and 5-0 R(0) HD lines at 1021.453 and 1042.847\,\AA, respectively.
Taking these results into account, an upper limit of 0.04\,\AA\ and 0.03\,\AA\
was subtracted from the \ovi\ equivalent width to HD\,116852 and
HD\,93840.  
HD\,116852 and HD\,93840 were the two only targets with measurable
absorptions of HD.
Following the data in Morton's (1991) list some metals may show here 
absorption lines, too. 
Since these particular elements have small intrinsic abundance
and since these lines are from excited states we can ignore 
any contribution. 
In most of our spectra, the weaker component of the \ovi\ doublet at
1037.61\,\AA\ is blended with the line of excited \cii\ at 1037.02\,\AA\ and
by two H$_2$ Lyman absorptions at 1037.15\,\AA\ and 1038.16\,\AA, both of
level J\,=\,1.  
Therefore, we used the weaker \ovi\ line only to verify the
result derived from the stronger component.

The velocity of the O\,{\sc vi} absorption lines usually agree well with
those of the C\,{\sc iv} lines as seen in IUE and HST spectra.
Apparently, the O\,{\sc vi} ions exist in gas well related with the gas 
containing C\,{\sc iv}. 
The complexity of the stellar background spectrum near the O\,{\sc vi}
absorption together with the intrinsic uncertainty in the IUE velocity scale
does not warrant a more detailed comparison at this time. 

\subsection{Galactic targets}

In all spectra of galactic stars which we have included to derive the \ovi\,
scaleheight, the H$_2$ Lyman P(3) line at 1031.19\,\AA\ was clearly separated
from the \ovi\ absorption.
The H$_2$ Lyman R(4) at 1032.35\,\AA\ blended with \ovi\ for several
targets. A two-component fit was used to separate the H$_2$ contribution to
the equivalent width for HD 77770, HD 93840 (Fig. 1b), HD 116852 and HDE
214930. 
In HD 93521 (Fig. 1a) and HD 49798 the R(4) absorption is clearly
separated.
In these cases a single gaussian was used to determine the equivalent width.

For HD 18100, HD 146813 and HDE 217505 no measurable absorptions from
neither H$_2$ Lyman P(3) nor H$_2$ Lyman R(4) was found between 990 and 
1120\,\AA, hence the contribution of any H$_2$ to the \ovi\ line can be
neglected in these cases.                           

\begin{figure}
 \resizebox{\hsize}{!}{\includegraphics{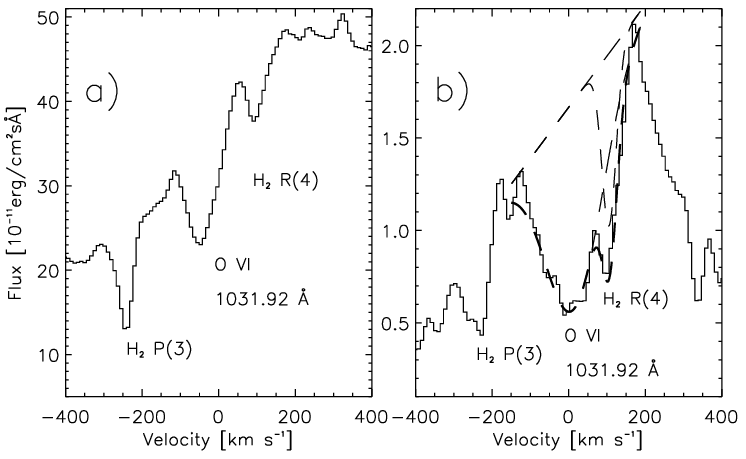}}
\caption[]{The \ovi\ absorption in the spectrum of a) the O9\,Vp star 
        HD 93521 and b) the B1\,Ik star HD 93840. \\ 
        a) The \ovi\ feature has a radial velocity of about $-$35 \kms. 
        Two neighbouring H$_2$ absorption lines are identified, too. \\ 
        b) Here the  H$_{2} $ R(4) line is not completely separated
        from \ovi} 
\end{figure}

\subsection{Magellanic Cloud targets}

Magellanic Cloud spectra have a poorer signal to noise ratio than 
those of the galactic stars. 
Yet, three velocity components can be clearly identified in each absorption. 
This means, that the H$_{2}$ lines mentioned above will blend 
(due to the complex velocity structure of the gas on these lines of sight) 
with the \ovi\ line, 
and decomposition may be problematic. 

\subsubsection {LMC stars}

In the spectra of HD 36402, HDE 269546, and Sk $-$67\degr 166 neither
measurable galactic H$_{2}$ absorption with rotational levels J $\gid$ 4 nor an
LMC component from any absorption with a level of J $\gid$ 3 are present.
Thus, for these three targets the zero velocity component was used to derive
the galactic \ovi\ column density.
We did not include the high negative velocity component we found in all
our extragalactic stars.
This feature seems to be composed of galactic H$_2$ Lyman P(3) absorption
and a second component not yet defined (Fig. 2). 

{\it LH 10:3120\,}:
In the \ovi\ line region de Boer \ea (1998) 
found H$_{2}$-absorption at LSR velocity as well as at +270 \kms.
The LMC component of H$_2$ P(3) 1031.19 \AA\ and the galactic
component of the H$_2$ R(4) 1032.35 \AA\ line blend with \ovi. 
Due to a lower S/N ratio, the separation of the galactic and LMC components
of these two H$_2$ absorptions was very inaccurate.
Therefore we just fitted the three velocity components mentioned above
(negative velocity, zero velocity and LMC component) to the absorption
profile.  
Referring to the results of de Boer \ea (1998) we subtracted 0.2 \AA\
in equivalent width from our zero velocity component.

\subsubsection{HD 5980 in the SMC} 
The SMC star HD 5980 shows a first positive velocity component at +147 \kms, 
the velocity of the SMC (Westerlund 1997).
In additon, we found at +300 \kms\ a velocity component 
clearly separated from the normal SMC absorption. 
It has also been seen in many other ions in IUE spectra of this star 
(Fitzpatrick \& Savage 1983) and
the authors suggested an expanding SNR in the foreground to the SMC star as a
possible origin of this feature.
The equivalent width of our derived zero velocity component leads to an
exceptionally small \ovi\ column density. 
The FWHM of 0.35\,\AA\ is also exceptionally small compared to the LMC
targets ($\sim$1.2\,\AA). 
It seems reasonable to suspect that galactic and SMC components are blended. 
However, given the uncertainties we decided not to include the \ovi\ 
information from the line of sight to HD 5980 in the scale height 
fit procedure.

\begin{figure}
 \resizebox{\hsize}{!}{\includegraphics{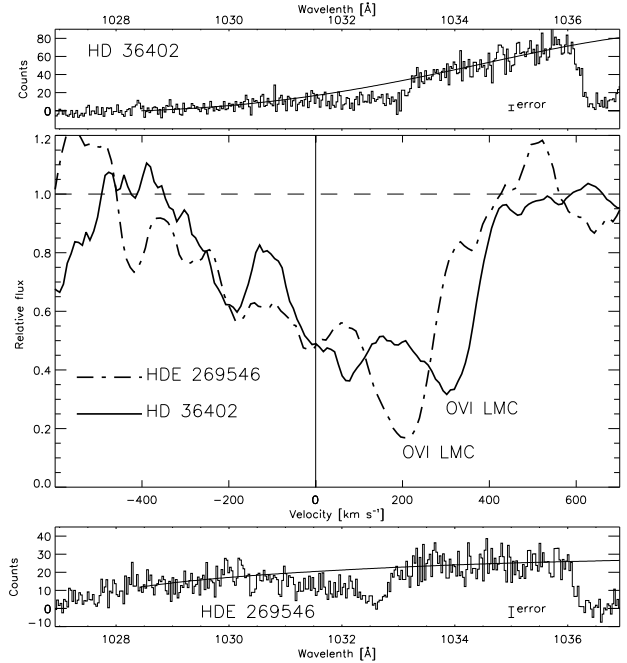}}
\caption[]{Spectra of HD 36402 and HDE 269546 centered at 1031.92 \AA.
           Relative flux is plotted against velocity shift. 
           The spectrum has been filtered with a de-noising algorithm basing
           on a  wavelet tranformation (Fligge \& Solanki 1997).
           The indicated Magellanic Cloud components of the \ovi\ absorption
           show almost the same velocity found for less ionized and
           neutral elements.
           Top and bottom panel show the original, unsmoothed spectra with
           their adopted continuum }
\end{figure}

\section{The spatial distribution of \ovi}

Our data allow to investigate the distribution of \ovi\ in the galactic
halo.
The absorption equivalent widths have been calculated from the result of
the fits.  
Assuming that absorption is optically thin, the column density can be 
calculated from 
\begin{equation}
N($\ovi$) [\mbox{cm}^{-2}] = 7.988 \times 10^{14} \times W_{\lambda} $[\AA]$
\end{equation}
\noindent
The column densities for all targets are given in Table 1. 

\begin{figure}
 \resizebox{\hsize}{!}{\includegraphics{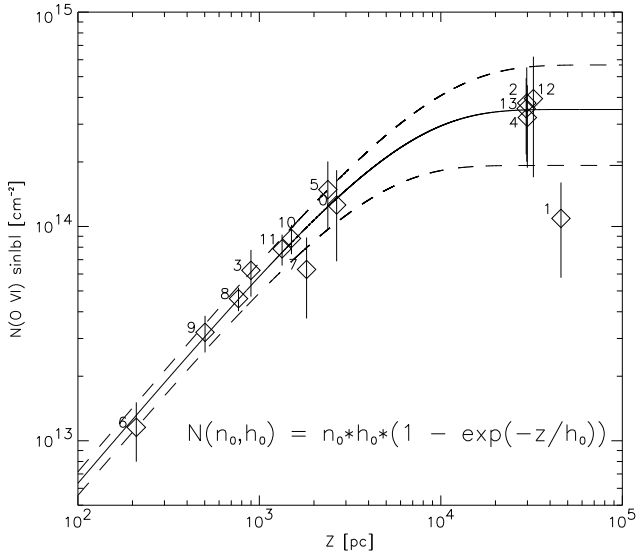}}
\caption[]{Plot of galactic $N$(\ovi)$\times \sin|b|$ versus $|z|$. 
        The solid line is the best-fit exponential, 
        the dashed line the $1\,\sigma$ deviation of the fit. 
        The derived values are $n_\mathrm{0} =
           2.07^{+0.26}_{-0.24} \times 10^{-8}$\,\cmc\ and $h_\mathrm{0} =
           5.50^{+2.37}_{-2.09}$\,kpc. 
           Each entry is marked with the target number (Table\,1)}
\end{figure}

If we assume a hydrostatic Galactic halo (corona) of the type postulated by
Spitzer  (1956), 
we have an exponential density distribution of  \ovi, 
described by the equation  
\begin{equation} 
 n = n_0 \times \exp(-z/h_0)
\end{equation}
\noindent
with the \ovi\ midplane density $n_0$ and scale height $h_0$. 
The projected column density is then given by 
\begin{equation}
N($\ovi$) \times \sin|b| = n_0 \times h_0 \times (1-\exp(-z/h_0)) 
\end{equation}
\noindent
Figure 4 shows the result we found when fitting equation (3) to our 
column densities.

Best values for the parameters $ n_\mathrm{0} $ and 
$h_\mathrm{0} $ resulting from our $N$(\ovi)$ \times \sin |b|$ fit are
$n_\mathrm{0} = 2.07^{+0.26}_{-0.24}\times10^{-8}$ \cmc\ and $h_\mathrm{0} =
5.50^{+2.37}_{-2.09}$ kpc.  
\noindent
Including the \ovi\ column densities from the literature, we find the data
point for 2C 273 to lie just above our 1$\sigma$ upper limit and those for
PKS\,2155-304 and NGC\,346 just below our 1$\sigma$ lower limit.
Therefore, inclusion of these data would not change our scaleheight
value in a significant way.

We should note here that $h_\mathrm{0}$ relies almost exclusively on the
LMC measurements. 
To set an approximate lower limit for the scaleheight we applied the fit
procedure to a dataset where the extragalactic stars are excluded. 
The result is $4.2^{+...}_{-2.8}$\,kpc, and points to a substantially
larger scaleheight of \ovi\ in the halo than previous measurements do.
The upper $1\,\sigma$ limit is $>$20\,kpc.

\section{Concluding remarks}
The \ovi\ column densities derived from our ORFEUS\,II data are far too
large to agree with the predictions of a photoionized model. 
These column densities clearly favour the hot halo concept as described by
Spitzer (1956) and  by Shapiro \& Field  (1976).  
The asymptotic column density of \ovi\ in Fig. 3 of 10$^{14.5}$ \scm\ can
be compared with that of \civ\ and \nv\ (Savage \ea 1997) of 10$^{14.1}$
and 10$^{13.4}$, respectiveley.  
If the \ovi, \nv\ and \civ\ were to coexist in space (the similar absorption
velocities of \ovi\ and \civ\ point to that) and ignoring the abundance of
the other ionic stages of these elements, the equivalent gas column 
density is $N$(H)\,$\simeq 10^{17.5}$\,cm$^{-2}$ (based on the
solar abundances $-3.4$ dex for C, $-4.1$ dex for N and $-3.2$ dex for O). 
Such identical equivalent column densities can be understood in a 
simple exponential pressure model, 
the small scaleheight for \ovi\ from previous measurements could not.
However, without knowledge of the real gas distribution in the halo 
it is not possible to relate our findings with the consequences of the 
interplay of ionisation and cooling in the halo. 

Due to its very high ionisation potential \ovi\ remains the most likeley
tracer of hot gas outflow from the Galaxy.
A direct measure of this outflow is not possible at 
the present time and will be subject of further research and 
the scientific goal for future Far Ultraviolet missions like 
ORFEUS\,III or FUSE.

\begin{acknowledgements}
ORFEUS could only be realized with the support of all our German and
American colleagues and collaborators.
We thank Ed Jenkins and the referee for helpful suggestions.
The ORFEUS project was supported by DARA grant WE3 OS 8501, WE2 QV 9304 and
NASA grant NAG5-696. PR is supported by DARA grant 50\,QV\,9701\,3
\end{acknowledgements}


\begin{thebibliography}{}

\bibitem[]{barnstedt}
Barnstedt J., Kappelmann N., Appenzeller I., et al., 1998, A\&AS submitted
\bibitem[1996]{chu}
Chu Y., Wakker B., Mac Low M., et al., 1994, AJ 108, 1696
\bibitem[1998]{deboer}
de Boer K.S., Richter P., Bomans D.J., et al., 1998, A\&A Letter, submitted
\bibitem[1998]{diplas}
Diplas A., Savage B. D., 1994, ApJS 93, 211 
\bibitem[1983]{fligge}
Fligge M., Solanki S. K., 1997, A\&AS 124, 579
\bibitem[1983]{fitzpatrick}
Fitzpatrick E.L., Savage B.D., 1983, ApJ 267, 93 
\bibitem[1991]{grewing}
Grewing M., Kr\"{a}mer G., Appenzeller I., et al., 1991, in: {\it Extreme
Ultraviolet Astronomy\/}, eds. R.F. Malina \& S. Bowyer, Pergamon Press,
p. 422 
\bibitem[1995]{hurwitz}
Hurwitz M., Bowyer S., 1996a, in: {\it
Astrophysics in the Extreme Ultraviolet\/}, eds. S.
Bowyer \& R.F. Malina, Kluwer; p. 601
\bibitem[1996]{hurwitz2}
Hurwitz M., Boywer S., 1996b, ApJ 465, 296
\bibitem[1995]{hurwitz3}
Hurwitz M., Bowyer S., Kudritzki R.-P., et al., 1995, ApJ 450, 149
\bibitem[1998]{hurwitz4}
Hurwitz M., Appenzeller I., Barnstedt J., et al., 1998, ApJ 500, L61
\bibitem[1978]{jenkins}
Jenkins E.B., 1978a, ApJ 219, 845
\bibitem[1978]{jenkins2}
Jenkins E.B., 1978b, ApJ 220, 107
\bibitem[1978]{jenkins3}
Jenkins E.B., 1978c, Comm. Astrophys. 7, 121
\bibitem[1978]{jenkins4}
Jenkins E.B., Wallerstein G., 1996, ApJ 462, 758
\bibitem[1990]{kraemer}
Kr\"{a}mer G., Barnstedt J., Eberhard N., et al., 1990, in: {\it Observatories
in Earth Orbit and Beyond\/}, ed. Y. Kondo, Kluwer, Ap.Sp.Sci.Lib.,
Vol. 166, 177  
\bibitem[1991]{morton}
Morton D., 1991, ApJS 77, 119
\bibitem[1952]{muench}
M\"{u}nch G., 1952, PASP 64, 312 
\bibitem[1957]{muench2}
M\"{u}nch G., 1957, ApJ 125, 42
\bibitem[1998]{richter}
Richter P., Widman H., de Boer K.S., et al., 1998, A\&A Letter, subm.
\bibitem[1973]{rogerson}
Rogerson J.B., Spitzer L., Drake J.F., et al., 1973, ApJ 181, L97
\bibitem[1997]{savage}
Savage B. D., Sembach K.R., Limin L., 1997, AJ 113, 2158
\bibitem[1992]{sembach}
Sembach K.R., Savage B.D., 1992, ApJS 83, 147
\bibitem[1976]{shapiro}
Shapiro P. R., Field G. B., 1976, ApJ 205, 762
\bibitem[1956]{Spitzer1}
Spitzer L.,1956, ApJ 124, 20
\bibitem[1956]{Spitzer2}
Spitzer L.,1996, ApJ 458, L 29
\bibitem[1997]{westerlund}
Westerlund B.E., 1997, {\it The Magellanic Clouds}, 
Cambridge University Press, Cambridge, New York 
\end{thebibliography}
\end{document}